\documentclass[9pt,twocolumn,twoside]{./osajnl}

\usepackage{textcomp}

\journal{./ao} 

\setboolean{shortarticle}{false} 

\newcommand\adeg{\mbox{$^\circ$}}%
\newcommand\lya{\mbox{Ly$\alpha$}}%
\newcommand\apj{\mbox{ApJ}}%
\newcommand\ao{\mbox{AO}}%
\newcommand\procspie{{Proc. SPIE}}%

\newcommand\alf{\mbox{AlF$_{3}$}}

\ifthenelse{\boolean{shortarticle}}{\colorlet{color2}{color2b}}{\colorlet{color2}{color2}} 

\title{Advanced Environmentally Resistant Lithium Fluoride Mirror
  Coatings for the Next-Generation of Broadband Space Observatories}

\author[1*]{Brian Fleming}
\author[2]{Manuel Quijada}
\author[3]{John Hennessy}
\author[1]{Arika Egan}
\author[2]{Javier Del Hoyo}
\author[2,4]{Brian A. Hicks}
\author[1]{James Wiley}
\author[1]{Nicholas Kruczek}
\author[5]{Nicholas Erickson}
\author[1]{Kevin France}

\affil[1]{Laboratory for Atmospheric and Space Physics, University of
  Colorado, 3665 Discovery Dr., Boulder, CO 80303}
\affil[2]{NASA Goddard Space Flight Center, 8800 Greenbelt Rd., Greenbelt, MD 20771}
\affil[3]{Jet Propulsion Laboratory, California Institute of
  Technology, 4800 Oak Grove Dr.,
  Pasadena, CA 91109}
\affil[4]{Department of Astronomy, University of Maryland, College
  Park, MD 20742}
\affil[5]{Center for Astrophysics and Space Astronomy, University of
  Colorado, 389 UCB, Boulder, CO 80309}

\affil[*]{Corresponding author: Brian.Fleming@Lasp.Colorado.edu}

\dates{Compiled \today}

\ociscodes{(160.4760) Materials, optical properties; (310.1515) Thin
  Films, protective; (310.1860) Thin Films, deposition and
  fabrication; (310.4165) Thin Films, multilayer design.}

\doi{\url{https://doi.org/10.1364/AO.56.009941}}

\begin{abstract}
Recent advances in the physical vapor deposition (PVD) of protective
fluoride films have raised the far ultraviolet (FUV: 912
-- 1600 \AA) reflectivity of aluminum-based mirrors closer to the
theoretical limit. The greatest gains, at more than 20$\%$, have come for lithium fluoride
protected aluminum (LiF+Al), which has the shortest wavelength cutoff
of any conventional overcoat. Despite the success of the NASA
$FUSE$ mission, the use of LiF-based optics is rare as LiF is hygroscopic and requires
handling procedures that can drive risk. With NASA now studying two large mission concepts for astronomy (LUVOIR and HabEx) that mandate throughput down to 1000 \AA, the development of LiF-based
coatings becomes crucial. This paper discusses
steps that are being taken to qualify these new enhanced LiF protected
aluminum (eLiF) mirror coatings for flight. In addition to quantifying the
hygroscopic degradation, we have developed a new method of protecting
eLiF with an ultrathin (10 -- 20 \AA) capping layer of a
non-hygroscopic material to increase durability. We report on the performance of eLiF-based optics and
assess the steps that need to be taken to qualify such coatings
for LUVOIR, HabEx, and other FUV-sensitive space missions.  
\end{abstract}

\setboolean{displaycopyright}{true}

\begin{document}

\maketitle
\thispagestyle{fancy}

\ifthenelse{\boolean{shortarticle}}{\ifthenelse{\boolean{singlecolumn}}{\abscontentformatted}{\abscontent}}{}

\section{Introduction}
The Lyman Ultraviolet (LUV, 912 $<$ $\lambda$ $<$ 1216 \AA) is one of
the richest bandpasses in astronomy, featuring important signatures
of molecular, atomic, and ionized gas tracing temperatures ranging from
10$^{2}$ -- 10$^{6}$ Kelvin. Poor mirror reflectivity has limited the capabilities of past space
observatories to explore this essential regime, making the LUV
the least explored bandpass in the UV-Vis-IR spectrum. To date only
approximately 5700 objects have been observed in the LUV at
sub-arcminute angular resolution, spanning less than 0.03$\%$ of the sky \cite{Fleming16}. This lack of data in the LUV
is a technical limitation, not one resulting from a deficit of
scientific interest \cite{Bowen09,Tumlinson12,France15,AURA15}.

The Decadal Survey Mission Concept Studies initiated by NASA in 2016
have identified four large space observatories for study in
support of the 2020 Astrophysics Decadal Survey. At least one of those
concept missions, the Large UV-Optical-IR Surveyor (LUVOIR) has
established LUV sensitivity to as short as 900 \AA\ as a
stretch goal \cite{Bolcar16}, with sensitivity to $\gtrsim$ 1000 \AA\
as a requirement \cite{AURA15,France16}. The Habitable Exoplanet
Imaging Mission (HabEx)
also contains a UV-sensitive spectrograph
\cite{HABEX16}. The choice of mirror coating is essential to
establishing the bandpass and throughput of these concept studies. 

Magnesium fluoride protected aluminum coatings (MgF$_{2}$+Al), such as
those used on the {\em Hubble Space Telescope} ($HST$), are $\gtrsim$ 80$\%$ reflective at wavelengths
$>$ 1150 \AA, but far less efficient ($\sim$ 15$\%$) at $\lambda$
$\leq$ 1100 \AA\ \cite{Osterman00}. As a result, sensitive FUV
spectroscopic instruments like the $HST$ Cosmic Origins Spectrograph
experience a precipitous three order of magnitude decline in effective
area between 1150 \AA\ and 1000 \AA\ \cite{Mccandliss10}. Recent advances in aluminum
fluoride (\alf ) deposition have driven this wavelength cutoff shorter
by $\sim$ 40 \AA\ on laboratory samples, but \alf +Al still lacks reflectivity
at important astrophysical lines such as \ion{H}{1} Lyman $\beta$ and
\ion{O}{6} \cite{Wilbrandt14,Bala15}. For observations deeper into the LUV, the state-of-the-art
coatings are silicon carbide (SiC), with a peak LUV reflectivity of
$\approx$ 40$\%$, and lithium fluoride protected aluminum (LiF+Al),
with a peak realized FUV reflectivity of R~$\approx$~67$\%$ for
$\lambda$ $\geq$ 1025 \AA\ \cite{Angel61,Ohl00}. Both SiC and LiF+Al were used
on the {\em Far-Ultraviolet Spectroscopic Explorer} ($FUSE$) \cite{Green94}, however
neither are suitable in their conventional forms for broadband
observatories as they significantly underperform MgF$_{2}$+Al from
$\lambda$ $>$ 1150 \AA\ $\rightarrow$ 1800 \AA. LiF is also a hygroscopic
material whose throughput will degrade with moderate exposure to
humidity, adding risk to large missions \cite{Angel61,Wilbrandt14}. 

\begin{figure}[!b]
\centering
\fbox{\includegraphics[width=\linewidth,trim={0.42in 0.15in 0.1in 0.25in},clip]{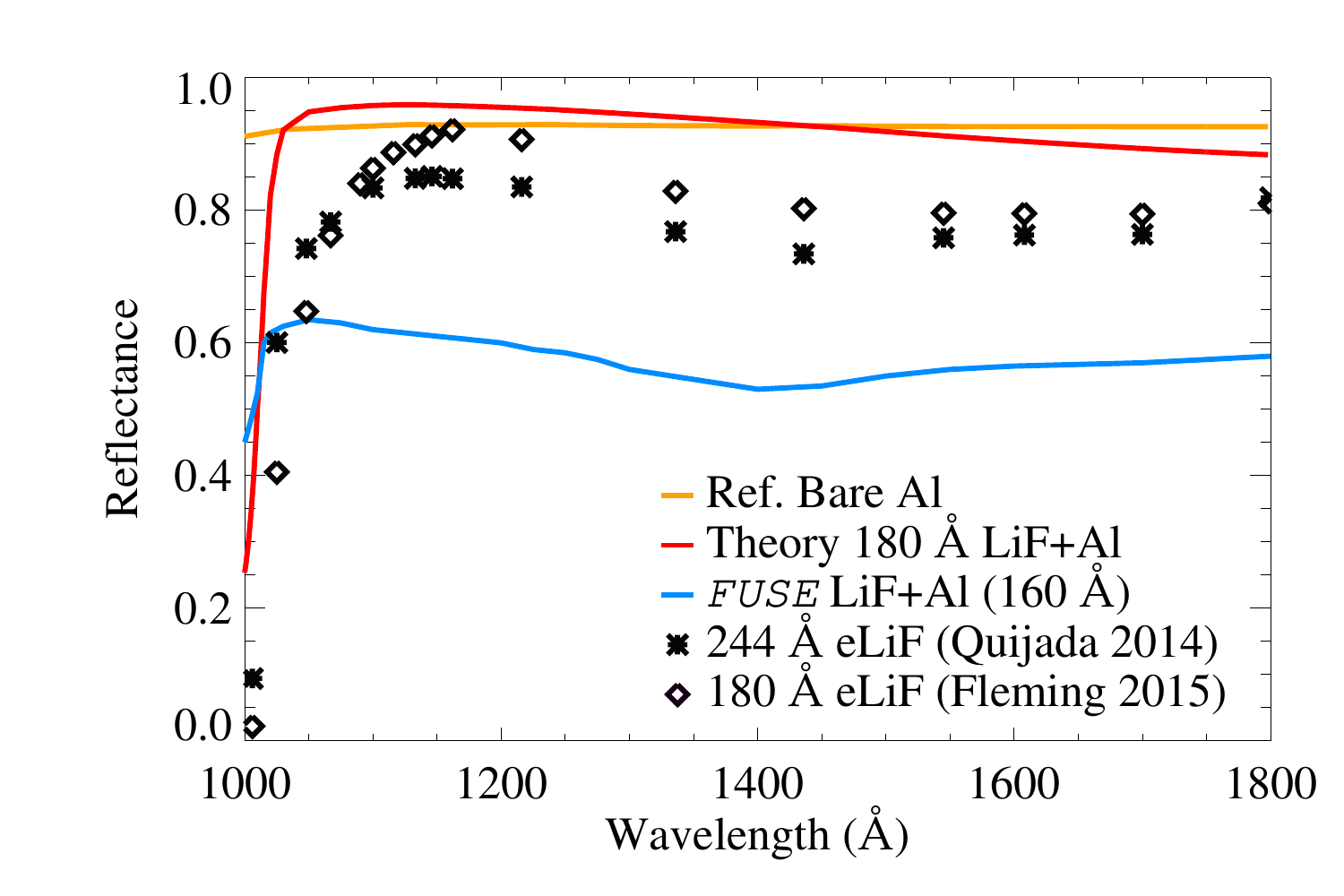}}
\caption{The theoretical reflectivity of (orange) bare aluminum and (red)
  LiF+Al, along with the measured reflectivity of (blue) $FUSE$ witness
  samples and (black) eLiF samples for which the LiF layer thicknesses
  were estimated as shown \cite{Quijada14,Fleming15}.}
\label{fig-coatings}
\end{figure}

There is significant room for improvement on both the net reflectivity
and hygroscopicity of LiF+Al films. Conventional physical vapor
deposition (PVD) techniques yield thin film coatings of LiF with high
surface roughness and low material packing densities, resulting in an
absorption coefficient in the FUV approximately four orders of
magnitude higher than that of bulk LiF crystal \cite{Adriaens71,Quijada12}. This results in a peak FUV reflectivity $\sim$ 30$\%$ less than the theoretical reflectivity
predicted by the optical constants of bulk LiF crystal and aluminum
\cite{Li76,Wilbrandt14}. Recent progress in new PVD
techniques developed at the NASA Goddard Space Flight Center (GSFC)
Thin Films Coating Laboratory (TFCL) have produced LiF+Al thin films
with higher
reflectivity in the LUV and lower surface roughness than has previously been obtained for
astronomical optics \cite{Quijada14}. These new
enhanced LiF+Al coatings (eLiF) promise to
reduce absorption and scatter losses,
improving the bandpass and throughput for next-generation ultraviolet space
observatories (Fig.~\ref{fig-coatings}) \cite{Fleming15,Fleming16}.


The hygroscopic degradation of LiF can be mitigated by maintaining a dry
integration and testing environment, typically with N$_{2}$ purging of
the instrument when not in vacuum, such as on the NASA Explorer
mission $FUSE$ \cite{Ohl00} and on numerous sounding rocket experiments
\cite{SliceSPIE,Hoadley16}. For large missions
such as LUVOIR, the risk associated with maintaining those measures on
dozens of meter-class optics may be prohibitive, and therefore a means
of permanently protecting the LiF from moisture exposure is
desirable. The current ``Architecture A'' version of the LUVOIR
Ultraviolet Multi-Object Spectrograph (LUMOS) requires a total of
seven reflections (four in the telescope, three in the instrument) to
image a spectrum \cite{LUMOS17}. While eLiF is currently the baseline mirror coating
used in throughput calculations for LUVOIR, hygroscopic degradation of
just 3$\%$ on each optical surface (the relative loss seen by the
$FUSE$ mission over instrument integration and testing \cite{Ohl00})
would result in a 22$\%$ relative loss of LUMOS
throughput. Conversely, similar gains in throughput can be realized from improved deposition
processes. 

In this paper, we present the first results of a study to
investigate overcoating eLiF mirrors with an ultrathin (10 -- 20 \AA)
layer of \alf\ deposited using atomic layer deposition (ALD). The
\alf\ transmissive bandpass does not extend to as short of wavelengths
as LiF, however at such a thin and
uniform layer the effects on the overall reflectivity may be minimal, and
\alf\ is believed to be more stable than LiF
\cite{Moore14,Hennessy16}. The capping of LiF with a protective overcoat
has been done before and has been shown to reduce hygroscopic
degradation \cite{Angel61,Wilbrandt14, Bala15}, however this is the first attempt
using ALD. The ALD process should produce a more uniform layer with
better layer thickness control than conventional evaporative processes
\cite{Johnson14,Hennessy15b}. These
protected eLiF samples were aged alongside bare eLiF and conventional
LiF+Al samples in a set of controlled humidors at the University of
Colorado (CU) Laboratory for Atmospheric and Space Physics (LASP) for
a period of approximately one year. The results of these aging
tests will establish the exact
degradation rate of eLiF and protected eLiF with humidity exposure,
providing a key parameter in
assessing risk for future UV-sensitive NASA missions. 

\section{Film Deposition Techniques}
We coated a batch of sixteen 50 $\times$ 75 mm fused silica slides of 1 mm
thickness with target eLiF and protected eLiF layer thicknesses of 700
\AA\ of aluminum, 160 -- 180 \AA\ of LiF, and 10 -- 20 \AA\ of \alf\ for the
protected eLiF. The LiF layer thickness target was chosen to optimize the LUV
reflectivity curve towards shorter
wavelengths around 1060 -- 1100 \AA. A thicker LiF layer would raise
the peak LUV reflectivity,
but also move the wavelength cutoff and reflectivity peak towards
longer wavelengths \cite{Angel61}. Depositions of eLiF were carried out at GSFC in March of
2016, with each of the 16 samples measured in the CU Square Tank
facility. Three eLiF samples were then sent to
JPL for the application of a protective overcoat of 10 \AA\ \alf\ on one
sample, and 20 \AA\ \alf\ on the other two. These samples were returned to CU for re-measurement before beginning the aging
experiment (\S\ref{initial}). All samples were stored in a dry N$_{2}$
purge box between deposition and aging to maintain the pristine
post-coating reflectivity. 

\subsection{eLiF Physical Vapor Deposition}

PVD is a common process for the deposition of thin films on mirrors
and other surfaces. Depositions are typically carried out using either
a sputtering target or via evaporation of the desired material, with
LiF+Al typically carried out via evaporation. The
GSFC TFCL has coated the optics of numerous astronomical, solar and
earth science missions and are a leader in film deposition
processes. 

The deposition chamber, in this case the 1 meter diameter chamber at
the TFCL \cite{Quijada12}, is evacuated to $\leq$ 10$^{-6}$ torr and
baked to outgas the materials. The chamber is left at high vacuum to cool overnight before starting the deposition process. Approximately 700 \AA\ of aluminum
is deposited at room temperature on the
optic surface by running a current through aluminum coated tungsten
filaments to create an aluminum vapor cloud. Shutters
over the filaments control the deposition thickness by limiting the
aluminum vapor exposure time. In a conventional LiF+Al deposition, 160
-- 180 \AA\ of LiF would then be applied approximately 3-5 seconds
later by passing a current through a molybdenum crucible
containing LiF powder and opening a separate shutter system. For small
optics the LiF crucible is centered beneath the substrates, while for
larger optics multiple crucibles can be spread around the chamber for
higher uniformity.

The deposition of eLiF requires heating the substrate to $\sim$ 250
\adeg C to lengthen
the freeze-out time of the LiF molecules when they condense on the
surface, increasing the resulting material packing density and
lowering the absorption coefficient. As soon as the aluminum
deposition cloud has cleared, a ``flash'' layer of $\sim$ 50 \AA\ of
conventional LiF is applied to partially protect the aluminum
from oxidization as the substrate is heated. The remainder of the desired LiF thickness is applied once
the substrate reaches the target temperature. Similar experiments have been
carried out demonstrating increased reflectivity from LiF+Al films
deposited on heated substrates, however it has never been done
for an astronomical optic \cite{Adriaens71, Hutcheson72}. 

The quality of LiF films is also dependent on
the deposition rate and chamber pressure, with higher deposition rates
generally resulting in higher quality films and less aluminum oxidization from the
residual gas in the chamber \cite{Hutcheson72,Wilbrandt14}. Too high of a deposition rate can result
in sputtering, however, therefore the rate is a controlled parameter \cite{Quijada14}. The depositions in this study
were carried out at approximately 4 nm/s, which was determined by
the GSFC TFCL to be an optimal rate for their system for conventional
LiF+Al. For a more complete
description of the eLiF process, see Quijada et al. 2014.

\subsection{AlF$_{3}$ Atomic Layer Deposition}
The protection of aluminum mirrors with fluoride films deposited by
ALD, which is a more controllable process than PVD, is a promising
field of research for future coatings
\cite{Hennessy15}. The protective fluorides typically
used on astronomical optics have all been deposited to surfaces via
ALD for other purposes \cite{Pilvi08,Lee16,Vos17}, including LiF \cite{Mantymaki13}. To date,
however, there has not yet been
a successful deposition of aluminum with a thin ALD overcoat that has
yielded higher reflectivities than eLiF
\cite{Moore14,Hennessy16}, largely due to aluminum oxidization in the
ALD chamber. While development efforts continue in the
field of ALD-only mirror coatings, it is important to continue development of more conventional processes that will result in
short term gains for new Explorer-class missions, and for initial LUVOIR and HabEx performance estimates. The process of depositing \alf\
via ALD onto optics was developed at the NASA Jet Propulsion
Laboratory (JPL) as part of one such complimentary study
into developing ALD-only mirror coatings for future missions \cite{Moore14,Hennessy16}. 

Three eLiF samples were packed in an ultra-high purity (UHP) N$_{2}$ environment to preserve the
pristine eLiF reflectivity and sent to JPL in April 2016 for protective \alf\
overcoating. These samples were not at risk of oxidization in the ALD
chamber as the aluminum was protected already by the LiF. Coating
thicknesses of approximately 10
\AA\ and 20 \AA\ were applied at a deposition temperature of 100 \adeg C to test the dependance of reflectivity
and hygroscopicity on the \alf\ layer
thickness. For a description of the ALD process for \alf\
deposition, see Hennessy et al. 2016. 

\section{\lowercase{e}LiF and Protected \lowercase{e}LiF Initial
  Reflectivity}\label{initial}

Unprotected eLiF coating reflectivities were measured for one sample at the GSFC TFCL
after coating, and for all samples in the CU Square Tank chamber
\cite{Fleming15} on April 24th and 27th, 2016 at eight discrete
hydrogen and argon spectral lines in the 973 -- 1608 \AA\ FUV
bandpass. 

\begin{figure}[!b]
\centering
\fbox{\includegraphics[width=\linewidth,trim={1.12in 0.85in 0.3in 0.35in},clip]{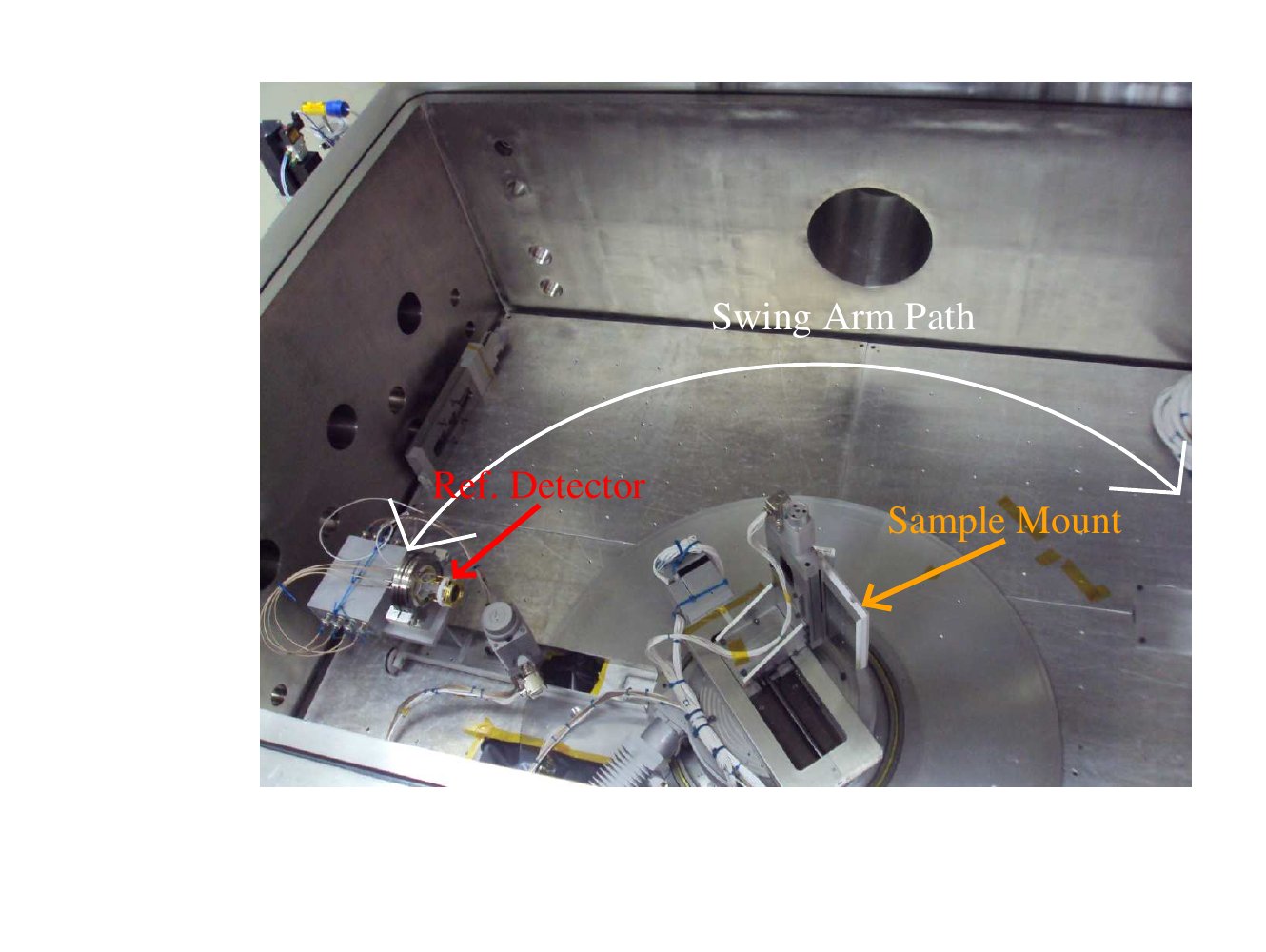}}
\caption{An annotated photo of the inside of the CU Square Tank. Samples are mounted at the axis of the swing arm stage.}
\label{fig-squaretank}
\end{figure}

The Square Tank FUV detector is a bare Quantar microchannel plate (MCP)
mounted onto a swing-arm stage with an arm length of approximately one meter. The samples are mounted at the swing arm pivot point
such that the MCP is always facing the sample mount. The mount is
attached to a linear vacuum stage which in turn is mounted on a rotation
stage - allowing the samples to be moved in and out of the beam, and
for control of the reflection angle. For all measurements the samples were held at a 7.5\adeg\
angle relative to the incident beam, resulting in a 15\adeg\ reflection angle. 

The light source is a hollow cathode arc lamp \cite{Fastie75} attached
to a Princeton/Acton VM502 monochromator with a collimating optic at
the monochromator output. The collimated, monochromatic beam is
directed into the Square Tank such that it passes over the pivot point
for both the swing arm and the rotation stage. The experimental setup
is presented with annotations in Figure~\ref{fig-squaretank}.

\begin{figure}[t]
\centering
\fbox{\includegraphics[width=\linewidth,trim={0.45in 0.15in 0.1in 0.25in},clip]{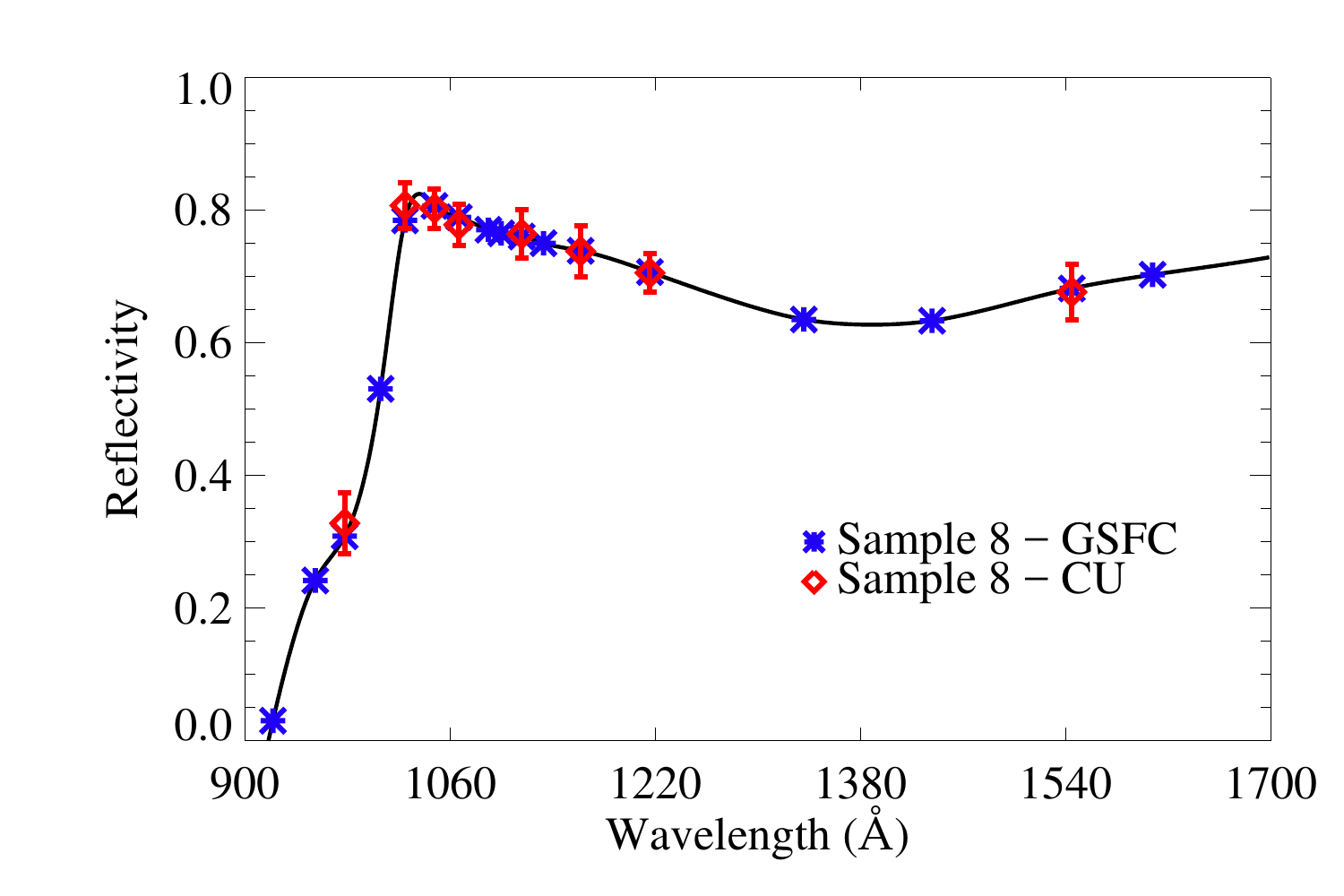}}
\caption{A comparison of the FUV reflectivity of Sample 8 as
  measured on 03/11/2016 at GSFC versus a measurement on 04/24/2016 in
  the CU Square Tank. The samples were stored in a dry N$_{2}$
  environment in the time between measurements and therefore no
  degradation was expected.}
\label{fig-compare}
\end{figure}

An incident light measurement is taken with the MCP in the 0\adeg\
position and with the samples shifted out of the beam path. The beam
intensity is controlled by an exit slit on the monochromator and set
to $\approx $ 4000 counts per second for each wavelength. A
standardized incident count rate reduces uncertainties due to MCP dead
time between wavelengths, as well as over the course of the experiment as
samples are aged. Two 10
second measurements are taken, and then a shutter is closed that
blocks the beam in order to measure
the system dark level (typically less than 5$\%$ of the incident intensity). The swing arm is then moved to 165\adeg\ and the
samples shifted back into the beam using the linear and rotation stages. The beam is
positioned so that it is incident on the same region of the MCP for
all measurements. Two reflection measurements are taken for each
sample, with two dark measurements in-between, and then the system is
returned to incident mode for a second set of incident
measurements. The entire procedure is repeated if the incident level is found to vary by more than a
few percent between the initial and final measurements for each
wavelength, mitigating the effects of variability in the light source. There was a
close agreement between the reflectivity determined at GSFC and the
reflectivity determined at CU for the only sample measured at both
locations (Sample
8, Fig.~\ref{fig-compare}). 

\subsection{Reflectivity Pre- and Post-ALD}
\begin{figure}[!b]
\centering
\fbox{\includegraphics[width=\linewidth,trim={1.62in 0.65in 0.55in 0.95in},clip]{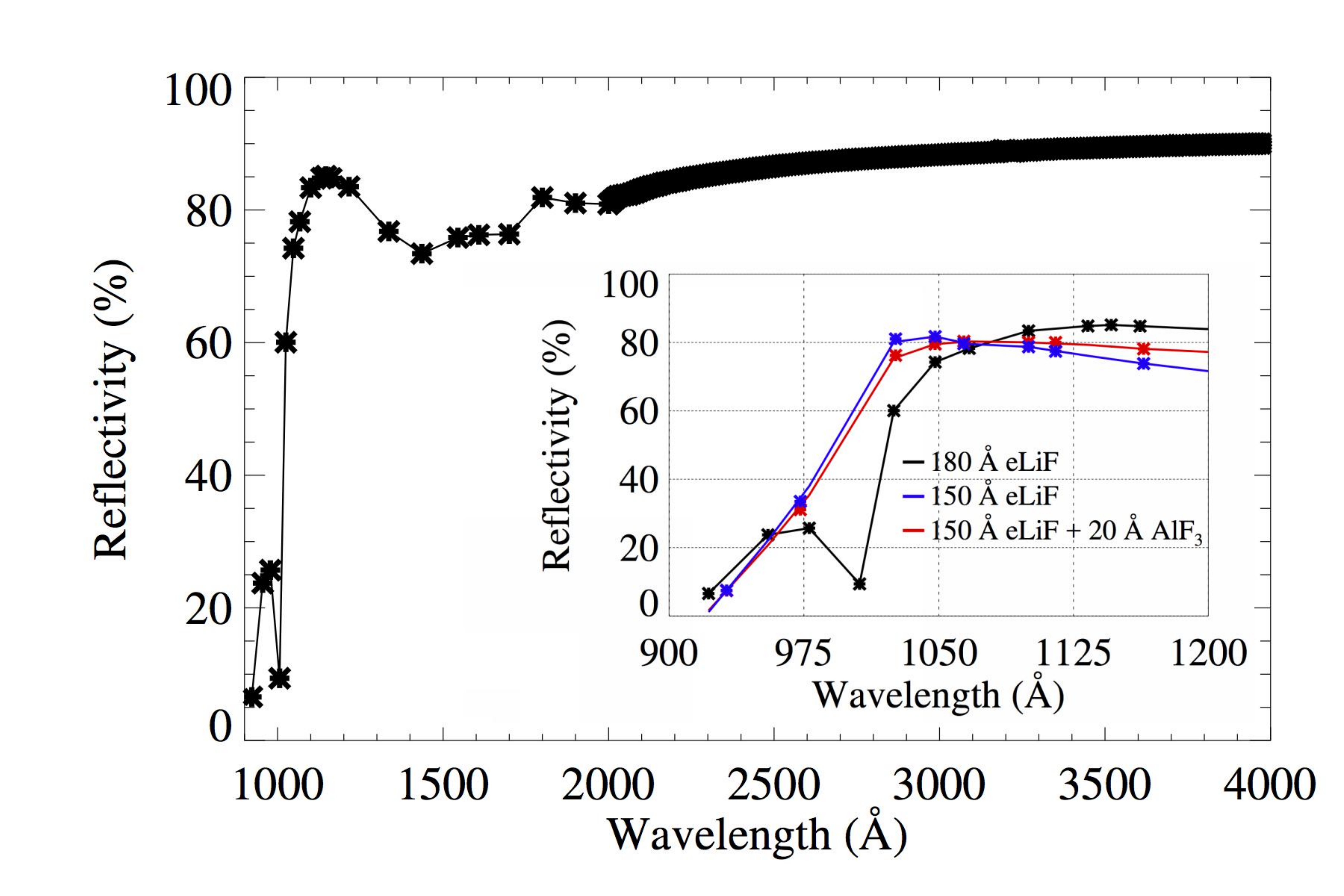}}
\caption{A calibration eLiF sample with an approximately 180 \AA\
  LiF layer thickness with an inset of the LUV region of the curve and the pre- and post-ALD reflectivity of Sample 6.}
\label{fig-prepost}
\end{figure}

The initial reflectivities of the pristine eLiF optics were higher at
\ion{H}{1} Ly$\beta$ and lower at \ion{H}{1} Ly$\alpha$ than
they were for a calibration sample
deposited August 2015, which is presented in
Figures~\ref{fig-coatings} ~and~\ref{fig-prepost} and was the target
we attempted to replicate for this experiment \cite{Fleming15}. This
peaking of the
reflectivity curve at shorter wavelengths is indicative of a LiF layer
thinner than the optimal 1/4 of the desired peak reflectivity
wavelength ($\sim$ 1100 \AA) within the dielectric. These
samples feature a peak reflectivity around \ion{H}{1} Lyman beta
($\lambda$ = 1026 \AA), with destructive interference of the
phase-shifted light reflecting off of the multilayer interfaces causing a drop in reflectivity for the rest of the
FUV bandpass, centered around 1380 \AA\ for Sample 8
(Fig.~\ref{fig-compare}). Assuming the index of refraction of the LiF
thin film is identical to that of bulk crystal \cite{Li76}, this corresponds to a
layer thickness of closer to 150 \AA\ for these samples, rather than
the target of 180 \AA.  

\begin{table}[!t]
\centering
\caption{\bf Reflectivities Pre- and Post-\alf\ Deposition}
\begin{tabular}{l|cc|cc|cc}
\hline
S$\#$ & \multicolumn{2}{c|}{Ly$\beta$ 1026 \AA} &
\multicolumn{2}{c|}{\ion{Ar}{1} 1067 \AA} & \multicolumn{2}{c}{\lya\ 1216 \AA} \\

& R$_{eLiF}$ & R$_{+AlF_{3}}$ & R$_{eLiF}$ &
R$_{+AlF_{3}}$ & R$_{eLiF}$ & R$_{+AlF_{3}}$ \\

\hline
S2 & 0.76 & ----- & 0.75 & ----- & 0.69 & ----- \\
S5$^{1}$ & 0.77 & {\bf 0.79} & 0.77 & {\bf 0.77} & 0.65 &
{\bf 0.73}  \\
S6$^{2}$ & 0.81 & {\bf 0.76} & 0.80 & {\bf 0.80} & 0.71 &
{\bf 0.77}  \\
S7 & 0.77 & ----- & 0.78 & ----- & 0.72 & ----- \\
S8$^{2}$ & 0.81 & {\bf 0.78} & 0.78 & {\bf 0.79} & 0.70 &
{\bf 0.74}  \\
S9 & 0.77 & ----- & 0.72 & ----- & 0.64 & ----- \\
C1$^{3}$ & 0.65 & ----- & 0.70 & ----- & 0.73 & ----- \\
C2$^{3}$ & 0.66 & ----- & 0.70 & ----- & 0.73 & ----- \\
\hline
\end{tabular}
\\
\raggedright
~~~~$^{1}$ eLiF  + $\sim$ 10 \AA\ \alf \\
~~~~$^{2}$ eLiF  + $\sim$ 20 \AA\ \alf \\
~~~~$^{3}$ Conventional LiF+Al control sample \\
  \label{tbl-prepost}
\end{table}

Six samples were selected for this initial aging study, with two
conventional LiF+Al samples coated in November 2016 added to the study
as
control samples (C1 and C2). Peak
reflectivity of the samples varied by $\sim$ 5$\%$ due to
coating thickness non-uniformities inherent to the 1m TFCL deposition chamber. Of these six,
three were selected for protective \alf\ overcoats at JPL such that
there would be a mix of estimated eLiF layer thicknesses. Table~\ref{tbl-prepost} shows the initial reflectivity of each
sample at \ion{H}{1} Ly$\beta$, \ion{Ar}{1} $\lambda$ 1067 \AA, and
\ion{H}{1} \lya, as well as the post-ALD reflectivity of the three protected eLiF
samples. We found a maximum reduction in reflectivity of less than 5$\%$ at
the shortest wavelength (Ly$\beta$) near the LiF cutoff, and an
increase in reflectivity at
\lya\ by as much as 8$\%$. These gains are likely a
result of the thickening of the total dielectric layer shifting the
peak reflectivity to longer wavelengths. It is not clear what the magnitude of the reflectivity
change due to the \alf\ layer would be
on an optimized eLiF sample, however it is clear that the
absorption at the short wavelength end will be on the order of only a
few percent. This is despite the fact that both Ly$\beta$ and
\ion{Ar}{1} $\lambda$1067 \AA\ are shortwards of the \alf\
cutoff. Previous studies that have employed a capping layer deposited
via PVD or other evaporative processes showed larger reflective losses
on the order of 10$\%$ relative to bare LiF+Al \cite{Angel61}. This first
test of ALD deposited thin protective films over environmentally
sensitive mirror coatings is therefore successful in maintaining an
eLiF-like reflectivity curve (Fig.~\ref{fig-prepost}).

\subsection{Uniformity of ALD \alf\ Layer}\label{uniform}

\begin{figure}[!t]
\centering
\fbox{\includegraphics[width=\linewidth,trim={0.2in 0.65in 1.3in
    0.3in},clip]{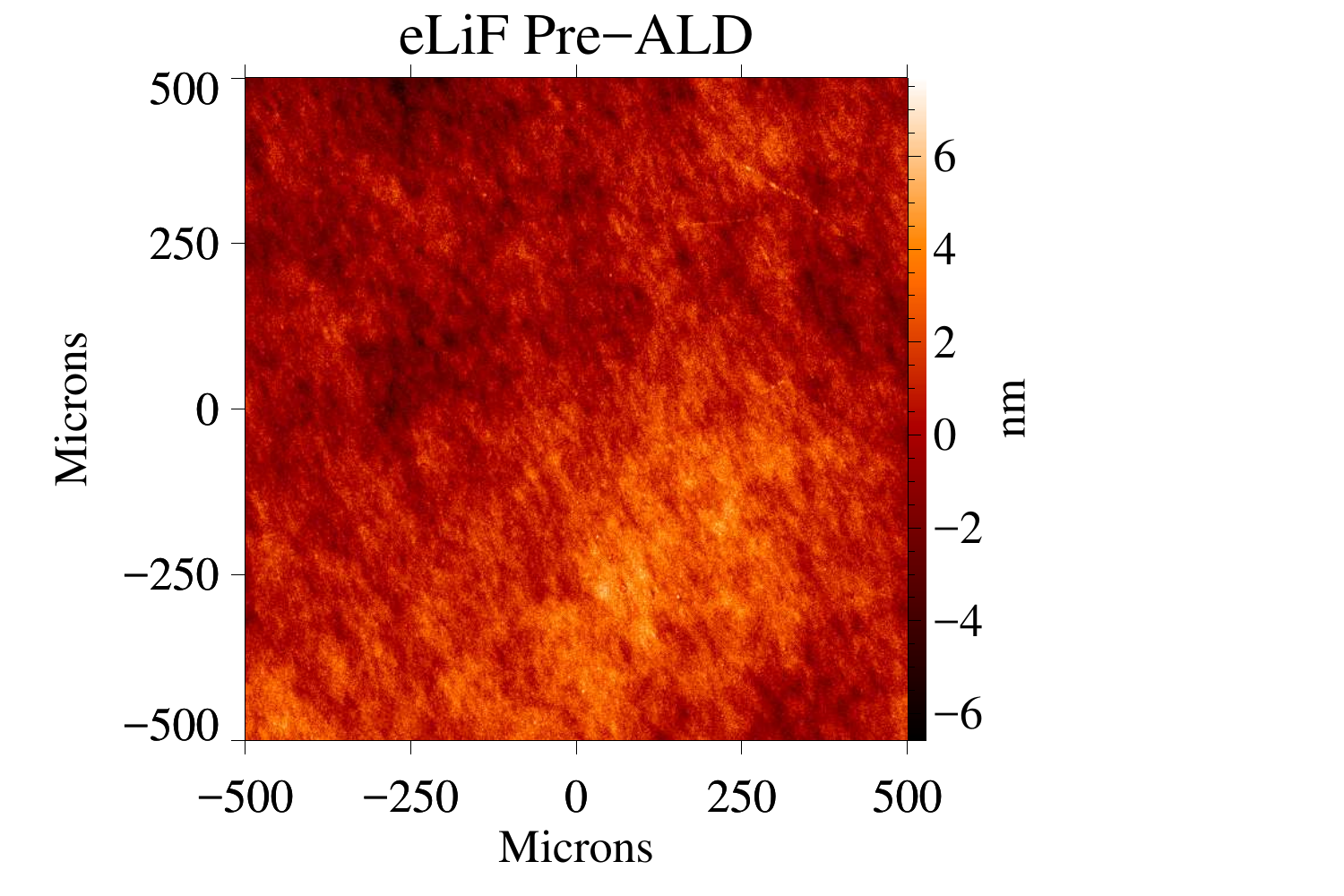}}\\
\fbox{\includegraphics[width=\linewidth,trim={0.2in 0.1in 1.3in 0.3in},clip]{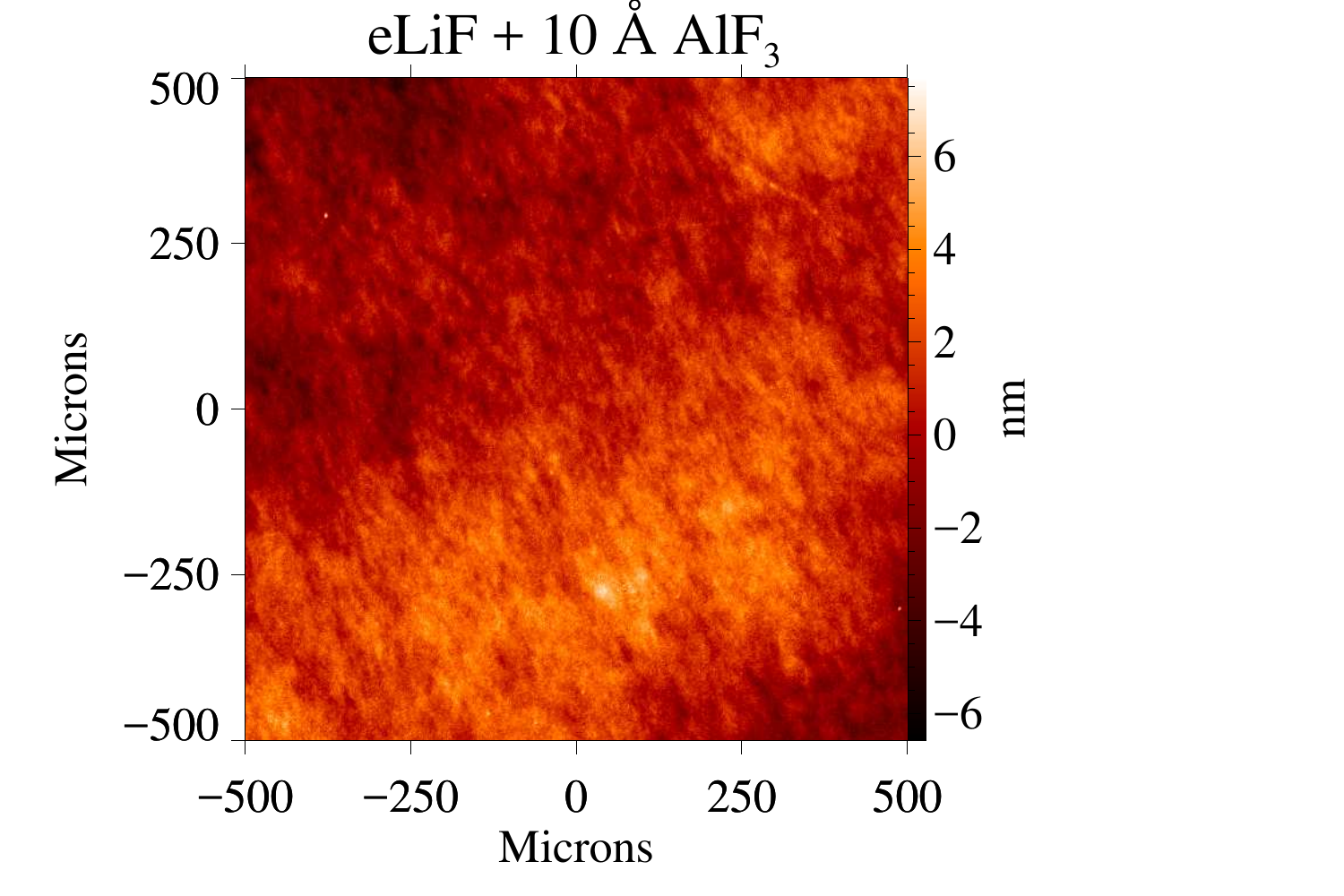}}
\caption{(Top) eLiF sample with $\sim$ 150 \AA\
of LiF on top of $\sim$ 700 \AA\ of Aluminum. (Bottom) The same region
after deposition of $\sim$ 10 \AA\ of \alf\ via ALD. Artifacts on the
optic surface were used to align these images.}
\label{fig-profiles}
\end{figure}

ALD produces highly
uniform layers, which should not only provide a greater material packing
density than evaporative processes, but also be mostly free of pinholes or other pathways for water
vapor to contact the LiF. \alf\ films deposited via this
process have been measured
at JPL to have approximately 1$\%$ uniformity over similar sized
wafers to those used in this study \cite{Hennessy15b}, however these
measurements were of samples without underlying PVD film(s). 

eLiF and protected eLiF samples were measured before and after \alf\
deposition using a Wyko NT2000 profilometer at the CU Keck Optics
Metrology laboratory (Fig.~\ref{fig-profiles}) to assess whether the
application of \alf\ via ALD over a PVD surface has a drastic effect on the surface morphology of the optic. This data incorporates the surface roughness of
the fused silica slides ($\sim$ 1$\lambda$), the $\sim$ 700 \AA\ of
aluminum, the $\sim$ 150 \AA\ of LiF, and the 10 -- 20 \AA\ ALD layer; therefore it
is unlikely that this thin capping layer will have any appreciable
effect on the profilometry unless the \alf\ does not adhere well to
the LiF and clusters or pools on the surface.
We find an RMS surface roughness of 1.6
nm in the sample pre-ALD, and 1.7 nm post-ALD, which is consistant with
our expectations of little to no change. 

Surface features and artifacts apparent in the profilometry data of the unprotected
eLiF samples are still apparent after the addition of the ALD
layer, which confirms that the measurement regions in Figure~\ref{fig-profiles}
are identical. There are variations in the relative heights
and depths between features at these scales on the order of $\pm$ 1-2 nm before and after the deposition of the
\alf, however it is not conclusive whether these changes are due to
the ALD process or simply instrumental variability between
measurements, which were taken three weeks apart. We conclude that
the general topology of the samples is preserved, and that the use of
ALD to protect eLiF optics does not greatly alter the
surface roughness.

\section{Humidity Sensitivity}\label{humidity}
\begin{figure}[!t]
\centering
\fbox{\includegraphics[width=\linewidth,trim={1.12in 0.85in 0.3in 0.35in},clip]{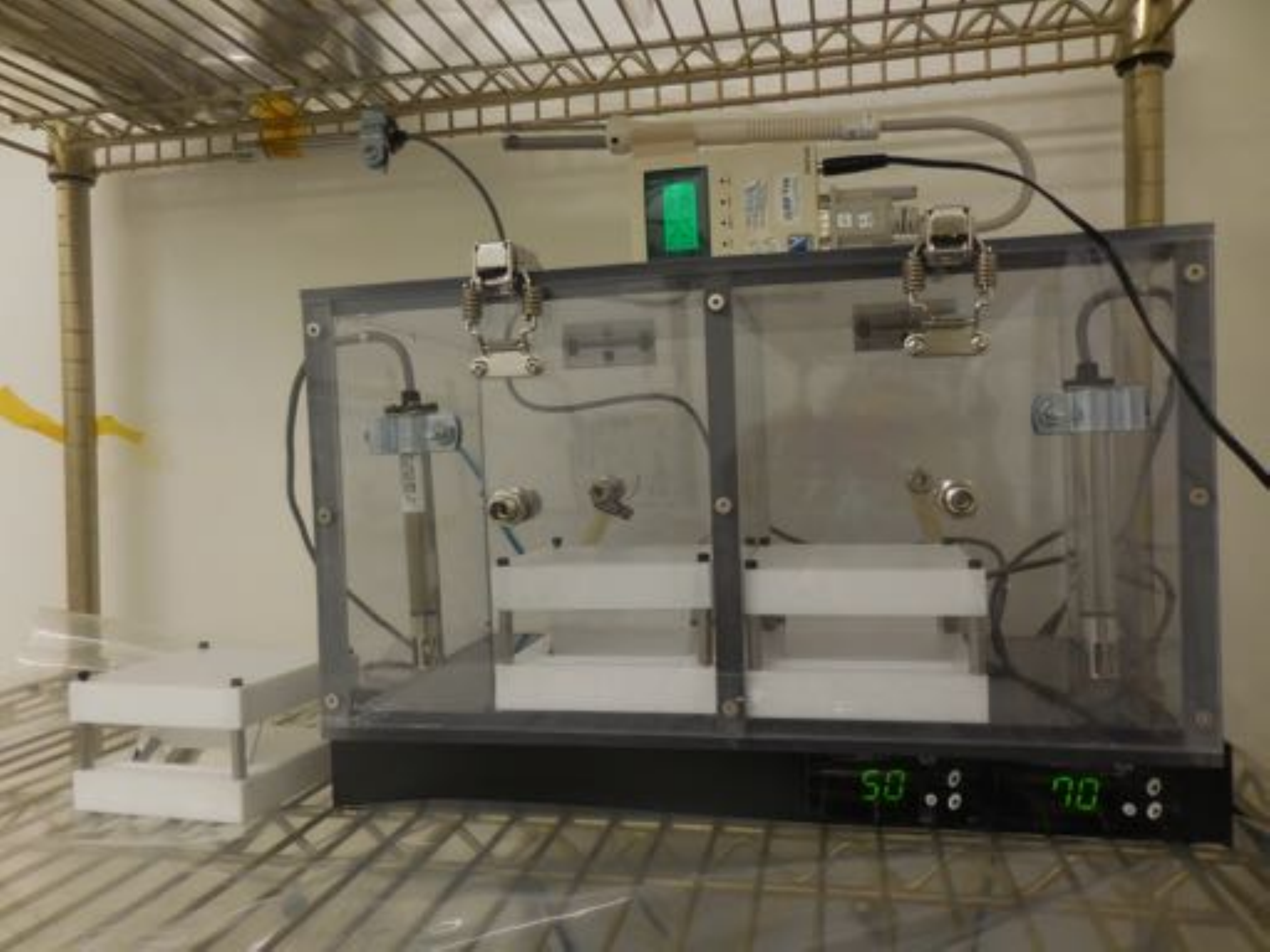}}
\caption{Custom humidors at fixed 50$\%$ and 70$\%$ RH in a CU clean
  room facility.}
\label{fig-chambers}
\end{figure}

The six samples were aged for $\sim$ 1 year in a set of custom
humidity control chambers held to 50$\%$ and 70$\%$ $\pm$ 1$\%$ RH
(Fig.~\ref{fig-chambers}). The two
conventional LiF+Al control samples were added in November 2016
(Table~\ref{tbl-storage}). These RH levels were chosen as 50$\%$ is approximately the
maximum ambient RH of Colorado and Southern California, while 70$\%$
represents a typical ambient humidity in Maryland or Florida. These
values encompass the higher end of the range of ambient humidities seen in
places where UV instruments are fabricated or launched, simulating the
exposure of the optics if no protective measures were taken during an
integration and testing phase. 

\begin{table}[!b]
\centering
\caption{\bf Sample Aging Conditions}
\begin{tabular}{l|ccc}
\hline
S$\#$ & Coating & RH & Date Started\\

\hline
S2 & eLiF & 70$\%$ & April 27, 2016 \\
S5 & eLiF+10\AA\ \alf & 70$\%$ & April 27, 2016 \\
S6 & eLiF+20\AA\ \alf & 70$\%$ & April 27, 2016  \\
S7 & eLiF & 50$\%$ & April 28, 2016 \\
S8 & eLiF+20\AA\ \alf & 50$\%$ & April 28, 2016  \\
S9 & eLiF & 70$\%$ & April 28, 2016 \\
C1 & LiF+Al & 70$\%$ & Nov. 24, 2016 \\
C2 & LiF+Al & 50$\%$ & Nov. 24, 2016 \\
\hline
\end{tabular}
\\
\raggedright
  \label{tbl-storage}
\end{table}

The humidors consist of a polycarbonate box fed by a balance of
humidified air sourced from ultrasonic humidifiers filled with deionized
water, and a soft ($\sim$ 0.1 psi) UHP N$_{2}$ purge \cite{Fleming15}. Omega Hx-71
humidity probes measure the RH in each box and feed an Omega RHCN-7000
AC humidity switch. When the
humidifiers are off, the low pressure N$_{2}$ purge lowers the
humidity in the chamber at a rate of approximately 0.5$\%$ RH per
minute, depending on the ambient humidity. AC power to the humidifiers is switched on when the humidity in the box falls below a
setpoint. The line pressure from the
humidifier overpowers the N$_{2}$ purge and raises the humidity in the
chamber. This keeps the humidity in the box to $\pm$ 1$\%$ of the
setpoint regardless of whether the ambient humidity is higher or lower
than the target. 

The Hx-71 probes were cross calibrated against a
secondary standard calibrated by Newport Corporation. Humidity values
were monitored by a LabJack U6 and logged every 30 seconds on a
Raspberry Pi. The samples were stored in a teflon holder with a lid and positioned such that the coated optic faced away from the gas inlet. This required
the humidified air to circulate and mix with the air in the chamber
before it diffused underneath the optic cover and exposed the coated optic
face, preventing direct exposure to the humidified air stream and
any water vapor particles that may be carried with it. The Hx-71 gauge
was positioned near the coated face of the samples (Fig.~\ref{fig-chambers}).  

\begin{figure}[!b]
\centering
\fbox{\includegraphics[width=\linewidth,trim={0.34in 0.20in 0.14in 0.55in},clip]{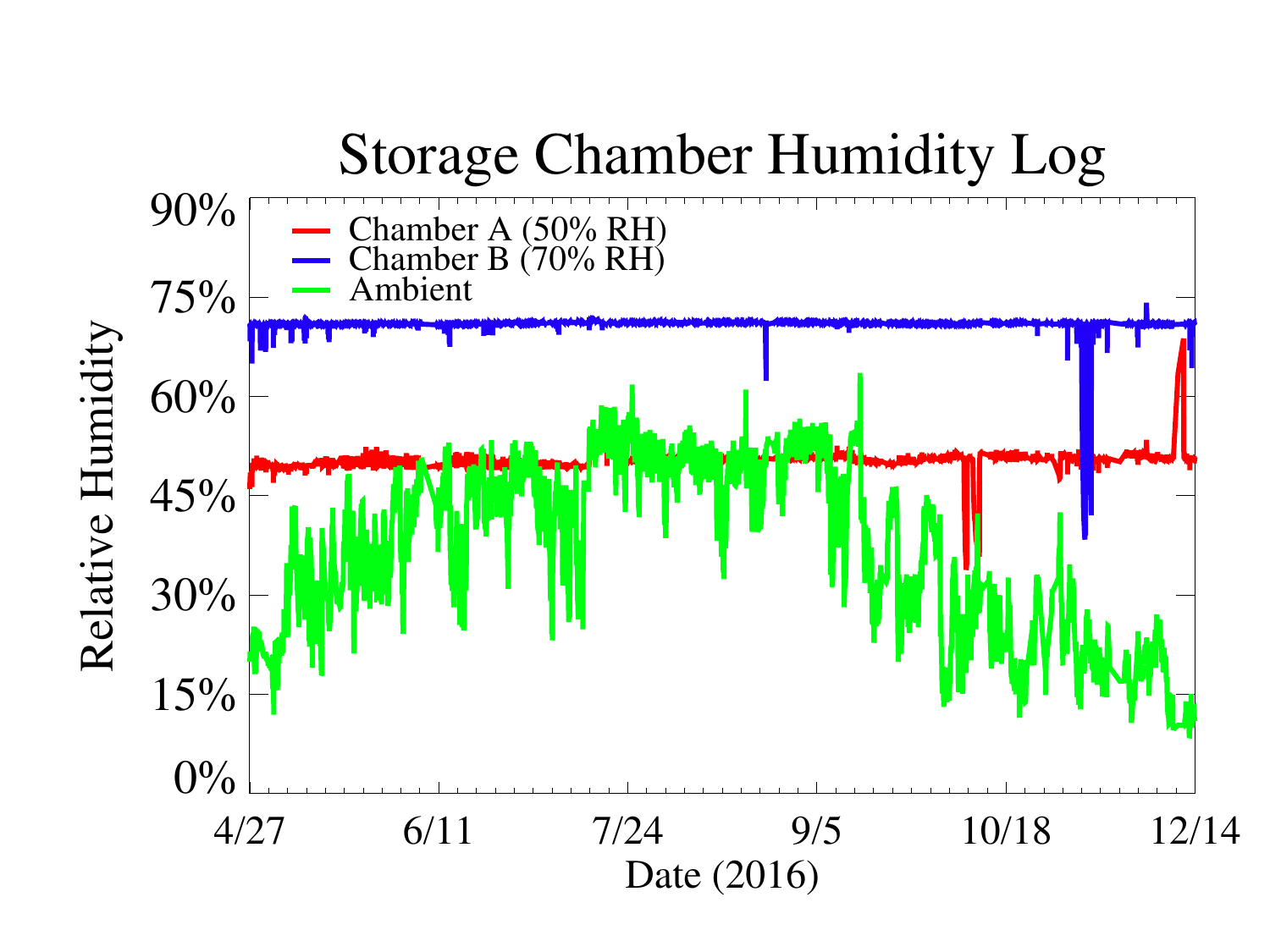}}
\caption{The recorded humidity of the humidor
  chambers and the ambient humidity. All humidity drops up until 10/30 are due to the chambers being opened to
  remove samples for measurement. The 70$\%$
  chamber dip at 10/30 is due to water runout, and the 50$\%$ chamber spike
  on 11/27 is due to a gauge failure. Humidity at this time was not
  accurately recorded.}
\label{fig-humidors}
\end{figure}

The humidor chambers were stored in a class 100,000 clean room with a filtered
air supply for 383 days from April 27, 2016 (April 28th for the 50$\%$
RH chamber) until May 15, 2017. There were no deviations from the RH
setpoint (save for when the chambers were opened to remove samples for
measurement) for the first 187 days of the experiment, after which the
70$\%$ chamber ran low of water for 4 days (temporarily lowering the RH). Then on
November 27, 2016, the humidity gauge on the 50$\%$ chamber failed, causing the
humidifier to lock into an ON state for $\sim$ 24 hours, saturating the
chamber with water vapor. This effectively ended the controlled
testing for the 50$\%$ chamber. From December 2016 to May 2017,
building construction disrupted the logging capability of
the chambers, forcing all further humidity monitoring to be done by
eye. With a few interruptions, both chambers remained at their
setpoint without exceeding it. Nevertheless, we consider the
experiment to be controlled only for the first 187 days, with subsequent data points suspect. The recorded humidity of each chamber for the
initial 232 days, as well as the ambient
humidity, is presented in Figure~\ref{fig-humidors}.

\subsection{Sample Degradation}\label{degrade}
\begin{figure}[!t]
\centering
\fbox{\includegraphics[width=\linewidth,trim={0.32in 0.15in 0.1in 0.55in},clip]{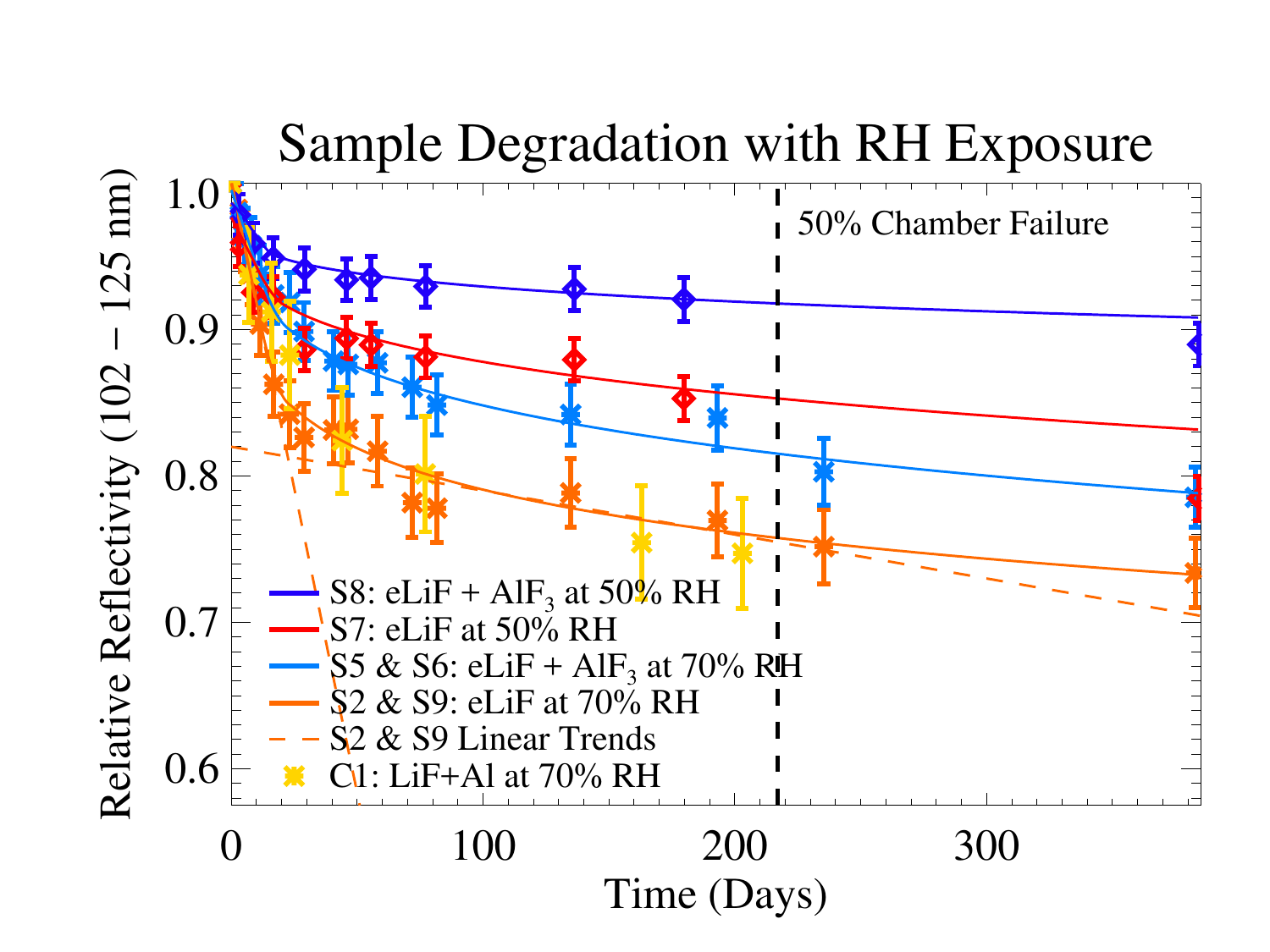}}
\caption{Relative reflectivity degradation as a function of
  time. Samples with similar coating prescriptions have been
  averaged. The date of the failure of the 50$\%$ chamber is marked. Linear
approximations of the degradation for one pairing are
overplotted. These approximations only consider data obtained before the chamber
failure.}
\label{fig-degrade}
\end{figure}

We found that the rate of degradation was similar for samples of the
same type stored in the same environments. At the end of the
experiment, S2 and S9 were within 3$\%$ of each other in relative
degradation. Likewise, S5 and S6 were within 2$\%$ of each
other, implying that there was no significant difference in protection provided by
10~\AA\ of \alf\ relative to 20 \AA. We therefore present
the average degradation of similar type samples from the 70$\%$ RH chamber in
Figure~\ref{fig-degrade}. We also include the individual 50$\%$ RH samples (S7
and S8) and the conventional LiF+Al control sample that was aged at
70$\%$ RH (C1). 

The samples all show signs of degradation averaged over the LUV regardless of whether
or not they are protected by \alf. The degradation is linear for the
first 10 -- 15 days of exposure, with the slope dependent
on both the level of RH exposure and the protection offered. The slope
shallows out over time before again following a trend that can be
approximated as linear, suggesting that the optics enter a semi-saturated state for a given maximum RH exposure. Unprotected eLiF
stored at 50$\%$ RH degraded in LUV reflectivity by approximately 13$\%$ (relative) over 6 months of
exposure, while unprotected eLiF stored at 70$\%$ RH degraded by 24$\%$
over the same time period. Both the control sample (C1) and the
unprotected eLiF samples aged in consistent manners, indicating that eLiF does not in itself offer more humidity
resistance than conventional LiF.

\begin{table}[b]
\centering
\caption{\bf Relative Degradation Over Six Months}
\begin{tabular}{l|ccc}
\hline
RH (\%) & Unprotected & Protected & $\%$ Difference \\
& Degradation & Degradation& \\

\hline
50$\%$ & 13.4$\%$ & 7.6$\%$ & 43.3$\%$ \\
70$\%$$^{1}$ & 23.6$\%$ &17.1$\%$ & 27.5$\%$ \\

\hline
\end{tabular}
\\
\raggedright
~~~~$^{1}$ Includes both eLiF  + 10 \AA\ \alf\ and eLiF + 20
\AA\ \alf \\

  \label{tbl-degrade}
\end{table}

The application of a thin layer of \alf\ applied by ALD offered a
significant level of protection in both chambers, with LUV degradations of
approximately 8$\%$ and 17$\%$ for the 50$\%$ and 70$\%$ samples, respectively.
This represents 43$\%$ and 28$\%$ less reflectivity loss for samples
stored in identical environments. These results are captured in
Table~\ref{tbl-degrade}.

\section{Discussion of Results}\label{results}
The rate of degradation of LiF-based optics under exposure to constant
humidity slows over time. For the purposes of easily estimating the risk to
astronomical optics during instrument integration and testing, we have
approximated the degradation rate using two linear functions: one over a $\sim$ 15 day
period and the other for an extended level of exposure of $\gtrsim$ 30 days,
with a curved transition in-between. We find that the degradation
slows by roughly an order of magnitude after the initial $\sim$ 15
day exposure, suggesting a partial saturation. The linear-fit
degradation rates per hour of exposure are presented in
Table~\ref{tbl-degraderates}.

\begin{table}[htbp]
\centering
\caption{\bf Relative Degradation Per Hour of Exposure}
\begin{tabular}{l|ccc}
\hline
Coating & RH (\%) & Rate (t $<$ 350h) & Rate (t $>$ 600h) \\

\hline
eLiF & 50$\%$ & 1.39 $\times$ 10$^{-4}$ & 1.06 $\times$ 10$^{-5}$ \\
eLiF+\alf & 50$\%$ & 0.76 $\times$ 10$^{-4}$ & 0.42 $\times$ 10$^{-5}$ \\
eLiF & 70$\%$ & 3.47 $\times$ 10$^{-4}$ & 1.25 $\times$ 10$^{-5}$ \\
eLiF+\alf $^{1}$& 70$\%$ & 2.13 $\times$ 10$^{-4}$ & 1.30 $\times$ 10$^{-5}$ \\

\hline
\end{tabular}
\\
\raggedright
~~~~$^{1}$ Includes both eLiF  + 10 \AA\ \alf\ and eLiF + 20 \AA\ \alf \\

  \label{tbl-degraderates}
\end{table}

\subsection{Potential Moisture Sensitivity of \alf}\label{activation}
\alf\ was selected for these initial ALD tests because it has the
closest bandgap to LiF of the other potential mirror coatings in use,
and because it is considered
hydrophilic, not hygroscopic. The incomplete protection of the samples to humidity exposure suggests
that it is possible water molecules penetrated the ALD layer to some
degree and degraded the underlying eLiF. It may be that \alf\ provides
complete resistance to RH exposure, but that the surface chemistry of
the ALD process is not compatible with eLiF and monolayers are not
evenly formed. The slower degradation of protected eLiF relative to
unprotected eLiF may
simply reflect the coverage percentage of the ALD layer. The ALD process should produce a more uniform and denser film
than PVD, however, and the similar
performance of both the 10 \AA\ and 20 \AA\ \alf\ overcoats aged at
70$\%$ RH suggest
that pinholes or incomplete coverage are unlikely to be the cause of the
sample degradation. More testing with a technique with higher
resolution than a profilometer,
such as atomic force microscopy (AFM), is necessary to prove this.  

\alf\ is a relatively new material for mirror coatings that has
never been used in an astronomical instrument, and this is the first controlled test of an \alf -based mirror at sustained
humidities $\gtrsim$ 50$\%$ \cite{Bala15,Hennessy16}. Given the paucity of experience with \alf ,
it is also possible that the incomplete protection of the eLiF optics is
due to unexpected moisture absorption by the \alf\ overcoat directly. It is known that
LiF stored in dry environments such as RH $<$ 25$\%$ desiccator boxes,
can last years with no evidence of degradation
\cite{Oliveira99,Hoadley16}, therefore our results suggest that there is an ``activation humidity'' under
which hygroscopic materials do not react significantly enough to
noticeably degrade. Previous tests of \alf +Al robustness may not
have breeched this threshold for an appreciable amount of time, and
therefore show no indication of significant degradation. 

That the hygroscopic degradation rate changes non-linearly with RH
exposure is not new and has been noted in previous experiments. In
Angel et al. 1961, LiF+Al was found to degrade by approximately 2$\%$ at
\ion{H}{1} Ly$\beta$ when stored for two months at laboratory
conditions of roughly 40$\%$ RH, but more rapidly for RH $\gtrsim$
50$\%$ \cite{Angel61}. They also suggested and tested overcoating LiF
with thin capping layers of approximately 15 \AA\ of MgF$_{2}$ applied
via conventional (not ALD) means. This lowered the initial reflectivity by $\sim$ 10$\%$. They found that
four days of aging at $\sim$ 50$\%$ RH led to a 22$\%$ degradation of the
unprotected LiF+Al optic, and 2$\%$ degradation for the protected
optic. 

If we were to assume that the degradation over six months as a function of
maximum RH exposure was a linear function, which is likely an
oversimplification given only two RH levels tested, we would derive
the following relationships for eLiF and \alf\ protected eLiF:

\begin{equation*}\
\begin{aligned}
& 1)  ~~ \Delta R = 0.51 (RH) - 12.1 ~~~ (eLiF) \\
& 2) ~~ \Delta R = 0.475 (RH) - 16.2 ~~~
(\alf\ Protected~
eLiF)
\end{aligned}
\end{equation*}

\noindent These relationships both would reach the point of 0
degradation in the 20 -- 40$\%$ RH range, with eLiF at $\approx$
23.5$\%$ (consistent with storage in an RH $<$ 25$\%$ desiccator box) 
and protected eLiF at $\approx$ 34$\%$. This would put protected eLiF
in the humidity range typical ESD (electrostatic discharge) clean rooms are held at (32 --
40$\%$), suggesting that a humidity controlled clean room and modest
purge procedures may be enough to preserve protected LiF-based optics. A wider range of humidities
would establish this relationship better and identify the maximum safe
humidity for these coatings. We have added two additional humidity
control chambers set to 25$\%$ and 40$\%$ RH for future tests.

\subsection{Resistance to Catastrophic Exposure}
The failure of the 50$\%$ storage chamber provides an opportunity to
assess the response of the coatings to short-term catastrophic
moisture exposure. There were three samples in the 50$\%$ chamber
when the Hx-71 gauge failed, setting the humidifier into an ON state
for approximately 24 hours. During this time, the chamber reached
humidities beyond the local dew point, resulting in water droplets
forming on all surfaces. In addition to S7 and S8, the chamber
contained a control sample (C2) that was added to the chamber in a pristine
state less than three days before the gauge failure. This provides a
serendipitous opportunity to compare the durability of eLiF and protected eLiF in the
semi-saturated state (S7 and S8, respectively) to unprotected LiF in a
pristine state (C2) when exposed to humidities higher than the local dew point.

Both semi-saturated samples exhibit a small but significant drop in
reflectivity as a result of direct exposure to highly humidified air,
with the protected eLiF sample dropping by about half the relative
reflectivity of the unprotected sample (Table~\ref{tbl-catastrophe}). This suggests that the
protected sample was more resistant to very high RH exposure and
condensation than unprotected eLiF. The
control sample drops by a much more significant 20$\%$
relative reflectivity, indicating that pristine optics (albeit
conventional LiF+Al, not eLiF) are more
sensitive to catastrophic exposure than optics in the semi-saturated state. C2 was aged in
the 50$\%$ chamber for the remainder of the experiment and was
measured periodically, during which time it degraded at a similar rate
as the semi-saturated unprotected eLiF did before the chamber failure (1.04 $\times$ 10$^{-5}$ per hour). The brief exposure to very
high humidity led to significant reflectivity losses, but once
humidity was restored it aged the same as the other semi-saturated unprotected
samples. 

\begin{table}[htbp]
\centering
\caption{\bf 1067 \AA\ Reflectivity After Condensation Exposure}
\begin{tabular}{l|cccc}
\hline
Sample & Coating & Ref. Prior & Ref. After & $\Delta$R \\

\hline
S7 & eLiF & 66.5$\%$ & 61.2$\%$ & 8.0$\%$ \\
S8 & eLiF+\alf & 71.8$\%$ & 69.4$\%$ & 3.3$\%$ \\
C2$^{1}$ & LiF+Al & 69.8$\%$ & 55.8$\%$ & 20.0$\%$ \\

\hline
\end{tabular}
\\
\raggedright
~~~~$^{1}$ Pristine - only aged 3 days before chamber failure \\

  \label{tbl-catastrophe}
\end{table}

\section{Conclusions and Future Work}
The protection of eLiF from hygroscopic degradation by applying an
ultrathin (10 -- 20 \AA ) overcoat of \alf\ via ALD was shown to
reduce the reflectivity degradation at FUV wavelengths, though it did
not provide complete hygroscopic immunity at the $\geq$ 50$\%$ RH
environments explored in this study. This matches similar
results from previous studies that have capped LiF with a protective
overcoat \cite{Angel61,Wilbrandt14}, though this is the first study to
apply the capping layer with ALD, and to age at controlled humidity
levels. The protected eLiF samples each degraded 28 -- 43$\%$ less
than unprotected eLiF when stored in identical environments over a six
month to one year timescale. Protected eLiF also showed greater
resistance to catastrophic exposure to water vapor on the timescale of
a few hours relative to unprotected eLiF. The application of the protective overcoat resulted in only a minimal
attenuation of the eLiF reflectivity curve beyond the \alf\ cutoff of
1070 \AA, which was possibly due more to the increase in the net
dielectric layer thickness than to attenuation by the \alf. 

Complete protection of the LiF layer coupled with further improvements in the
deposition process could make protected eLiF the standard coating for
all UV-sensitive observatories in the near future. Reflectivity peaks
of greater than 90$\%$ at \ion{H}{1} \lya\ have already been realized with
thicker coatings, matching the performance of MgF$_{2}$ at that
wavelength while still providing LUV sensitivity not possible with
other protected aluminum coatings \cite{Quijada14}. Additional
investment in the eLiF process optimization could lead to greater
gains, as eLiF
reflectivities remain 5 -- 10$\%$ below the theoretical maximum
estimated from bulk LiF crystal optical constants (Fig.~\ref{fig-coatings}).

We are currently preparing to expand this first ALD trial by protecting eLiF with MgF$_{2}$, rather than \alf. The
success of the \alf\ ALD process on maintaining the eLiF reflectivity
implies that similar success may be possible with MgF$_{2}$ - a
material with far more flight and laboratory heritage than \alf. We
have also obtained \alf\ + Al optic samples coated via a similar high temperature PVD process
at GSFC to determine the hygroscopic sensitivity, if any, of \alf\ in a
controlled setting. This test will enable us to determine whether the
degradation seen in our \alf-protected eLiF samples was due to
incomplete protection, or intrinsic moisture sensitivity in
\alf\ films. 

For this next
project phase we have constructed two additional humidity control
chambers that will operate at 25$\%$ and 40$\%$ RH to test for an
``activation humidity'' to fluoride degradation
(\S\ref{results}.\ref{activation}). If it can be shown that there is a humidity
floor below which hygroscopic degradation is not an issue, then it may
be possible to move away from the extreme protection measures utilized
successfully on $FUSE$ and other projects with LiF mirror
coatings. This will make it easier to estimate the cost and risk
associated with using hygroscopic optical coatings on future NASA
Explorer and large missions. We have also developed a new vacuum
measurement chamber designed specifically for high cadence, highly
repeatable reflectivity measurements for more efficient monitoring of
sample aging \cite{Wiley17}.  

Another advantage of eLiF is that unlike the ALD
process or other advanced conceptual coating strategies, unprotected eLiF can be
applied in the same GSFC deposition chambers that coated the $FUSE$ optics
and numerous other NASA missions. This enables a rapid path to flight
qualification and TRL advancement. In collaboration with the GSFC
TFCL, The CU/LASP ultraviolet sounding
rocket program is preparing to fabricate and launch the Sub-orbital
Imaging Spectrograph for Transition-region Irradiance from Nearby
Exoplanet host stars (SISTINE) sounding rocket payload, which will
serve as a flight test platform for eLiF-based optical
coatings \cite{Fleming16}. SISTINE will feature a 0.5 m primary mirror coated in
unprotected eLiF, as the JPL ALD chambers cannot currently accomodate
large optics, while the secondary and fold mirrors will be
overcoated at JPL with a protective layer of \alf\ or MgF$_{2}$ to demonstrate both eLiF
variants in a working instrument. Tests are ongoing at CU to determine
whether ion etched holographic gratings are compatible with high temperature
deposition processes. The coating of these flight optics will take place in early
2018 with a target launch date in 2019. SISTINE is designed to serve
as a platform for flight testing any new advanced LUV mirror coating
developed in the next few years, supporting rapid flight qualification
of this critical technology. 

Protected eLiF mirror coatings currently provide the highest total LUV
reflectivity from 1000 -- 1150 \AA\ with less humidity
sensitivity than unprotected eLiF. Reflectivities of $>$ 75$\%$
throughout the 1040 -- 4000 \AA\ bandpass ($>$ 85$\%$ for $\lambda$
$>$ 2500 \AA) have been demonstrated on
bare eLiF, with peak LUV reflectivities of $>$ 85$\%$. While this
first run of protected samples underperformed those previous results
due to an overly thin eLiF layer, protected eLiF is still the closest
of any existing mirror coating to meeting the technical requirements
of the LUVOIR project \cite{Bolcar16}. Further improvements in the eLiF deposition
process are underway at GSFC, while experiments with MgF$_{2}$
protected eLiF may offer even greater resilience, making protected eLiF the most advanced mirror coating for
LUV-sensitive space observatories currently available. 

\section{Funding Information}
This work was funded by a Nancy Grace Roman Technology Fellowship in
Astrophysics awarded by the National Aeronautics and Space
Administration (NASA) with grant number NNX15AF27G. The process
development to apply \alf\ via ALD was performed at the Jet Propulsion
Laboratory, California Institute of Technology, under a contract with
NASA. The authors would like to thank the anonymous reviewers for
providing helpful comments. 

\textcopyright 
2017 Optical Society of America. One print or electronic copy may be made for personal use only. Systematic reproduction and distribution, duplication of any material in this paper for a fee or for commercial purposes, or modifications of the content of this paper are prohibited.




\ifthenelse{\equal{\journalref}{ao}}{%
\clearpage
}{}

\end{document}